\documentclass[prb,twocolumn,showpacs,showkeys]{revtex4}
\usepackage{graphicx}
\usepackage{graphics}
\usepackage{dcolumn}
\usepackage{bm}
\newcommand{\beq}{\begin{equation}}
\newcommand{\eeq}{\end{equation}}
\newcommand{\beqnar}{\begin{eqnarray}}
\newcommand{\eeqnar}{\end{eqnarray}}
\newcommand{\bfig}{\begin{figure}}
\newcommand{\efig}{\end{figure}}

\begin{document}
\title{Nonlinear Transport through ultra-narrow Zigzag Graphene Nanoribbons: non-equilibrium charge and bond currents}

\author{Hosein Cheraghchi}

\affiliation{School of Physics, Damghan University, P. O. Box:
36715- 364, Damghan, IRAN } \email{cheraghchi@du.ac.ir}
\date{\today}
\newbox\absbox

\begin{abstract}
The electronic nonlinear transport through ultra narrow graphene
nanoribbons (sub-$10nm$) is studied. A stable region of negative
differential resistance (NDR) appears in the I-V characteristic
curve of {\it odd} zigzag graphene nanoribbons (ZGNRs) in both
positive and negative polarity. This NDR originates from a
transport gap inducing by a selection rule which blocks electron
transition between disconnecting energy bands of ZGNR. Based on
this transition rule, on/off ratio of the current increases
exponentially with the ribbon length up to $10^5$. In addition,
charging effects and also spatial distribution of bond currents
was studied by using non-equilibrium Green's function formalism
in the presence of e-e interaction at a mean field level. On the
other hand, we also performed an {\it ab initio} density
functional theory calculation of transmission through a
passivated graphene nanoribbon to demonstrate robustness of the
transport gap against hydrogen termination of the zigzag edges.

\end{abstract}


 \maketitle
\section{Introduction}
Graphene is a two-dimensional carbon which has been recently
explored \cite{novoselov}. Experiments in graphene-based devices
\cite{experiment} have shown the possibility of controlling their
electrical properties by the edge structure and also application
of an external gate voltage. Nowadays, by using a chemical
method\cite{Dai}, it is possible to fabricate field effect
transistor graphene nanoribbons (GNRs) with ultra narrow widths
and smooth edges which are possibly well defined zigzag or
armchair edge structures. A useful transport gap is opened in
sub-$10 nm$ GNRs at room temperature which results in a high
on/off current switching up to $10^6$. The origin of the
transport gap can be understood by two factors: suppression of
transport due to edge disorder leading to Anderson localization,
confinement along transverse direction\cite{han,Molitor}.
Additional to these factors, in ultra narrow zigzag graphene
nanoribbons (ZGNR), flow of current is also blocked by the
symmetric selection rules
\cite{cheraghchi1,grosso,beenaker,P-Ngraphene,duan,kurihara,Wakabayashi}.

Negative differential resistance \cite{esaki} in nanoelectronic
devices has been also observed in the metallic nanotube junctions
\cite{farajian} and in the case of potential barriers in $2D$
graphene sheets\cite{NDRgraphene}.
\bfig
\includegraphics[width=9 cm]{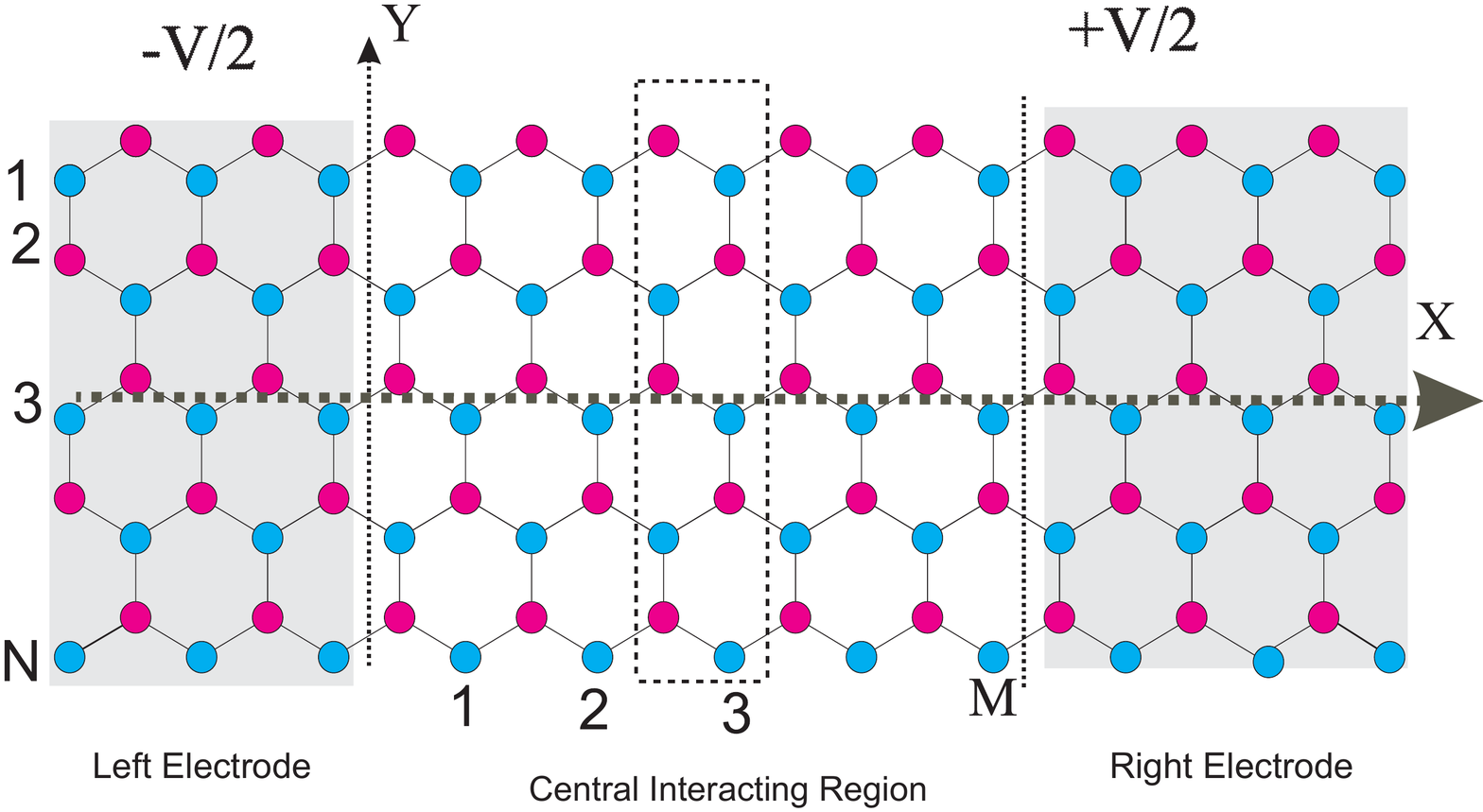}
 \caption{Odd zigzag graphene nanoribbon considered as a
central region attached to two electrodes. Transverse zigzag
carbon chains and unit cells inside the central region are
labeled by N and M.}\label{nanoribbon} \efig

In this paper, to shed light on the experimental work in
Ref.[\onlinecite{Dai}], we study non-linear transport in odd
zigzag graphene nanoribbons by using non-equilibrium Green's
function formalism (NEGF). It is shown that a stable NDR against
electrostatic interaction up to $10^5$ is appeared in ultra
narrow ZGNRs around $\pm 1t$ in both positive and negative
polarity, where $t$ is hopping integral between carbon atoms.
On-off ratio of this NDR increases exponentially with the ribbon
length. The NDR phenomenon occurs in far-from-equilibrium regime.

Although parity is not conserved in odd ZGNRs, current reduction
is induced by an other selection rule in which transition between
disconnecting band groups is forbidden (Cresti {\it et
al.}\cite{grosso}). The edges of graphene ribbons can also simply
absorb some chemical compounds\cite{kobayashi}. It is demonstrated
that this NDR is not much sensitive to the asymmetry of ribbons,
while transport gap in even ZGNRs which is based on the reflection
symmetry, is completely destroyed due to the edge disorder. The
above mentioned selection rule is valid for both even and odd
ZGNRs, while parity selective tunneling just belongs to even
ZGNRs. This transport gap is robust against hydrogen termination.
Additional to the model calculation, by using TranSIESTA
code\cite{siesta}, we performed an {\it ab initio} density
functional theory calculation of transmission through graphene
nanoribbon with hydrogen termination of the zigzag edges. Results
of first principle calculation confirms the transport gap
originating from transition rule between disconnected bands.

This NDR is stable against the presence of the electron-electron
(e-e) interaction at a mean-field approximation. It is concluded
from self-consistent charge and potential profiles so that
external potential is screened by charge redistribution around
the contact junctions. So potential profile deep inside the
ribbon remains flat. Furthermore, the e-e interaction increases
on-off ratio of the current. Moreover, in low and high applied
biases, we study spatial profile of local currents in the
presence of e-e interaction which contrasts with non-equilibrium
charge profile.

The paper is organized as the following sections. After an
introduction, we explain Hamiltonian and a short review of NEGF
formalism in section II. In section III, we present our results
containing NDR in I-V curve and charging effects in high applied
bias and spatial bond currents. Before to end the paper, we
present the results of {\it ab initio} calculated by using
Transiesta code. Finally, we conclude our results.
\section{Formalism}
The single electron Hamiltonian of the molecule is defined as:

\beq\begin{array}{r} H\{n\}=\sum_{i=1}^{2N \times M
}[\varepsilon_i+u^{ext}_{i} +\sum_{j=1}^{2N \times M}
V_{ij}\delta n_j] c^{\dagger}_{i}c_{i}
\\ +\sum_{<ij>}t (c^{\dagger}_{i}c_{j}+c_{i}c^{\dagger}_{j})
 \end{array} \label{hamiltonian}\eeq

where $c^{\dagger}_{i}$ and $c_{i}$ are the electron creation and
annihilation operators, respectively. $2N \times M$ and $t$ are
number of atomic sites and hopping energy between nearest
neighbor atoms. One $\pi$ orbital is considered per each site for
graphene as a planar 2D system. Without losing any generality, we
set onsite energies ($\varepsilon_i$) of all sites to be equal to
zero. All energies and voltages are in unit $t=2.7 eV$. The Fermi
energy of undoped graphene nanoribbon is at half-filling $E_F=0$
due to perfect electron-hole symmetry. By applying a source-drain
bias, the site energies are shifted by a linear potential
distribution along the molecule $u^{ext}_{i}$ which is the
solution of the Laplace equation. The applied bias $V$ is divided
symmetrically on the left and right electrodes as $-V/2$ and
$V/2$, respectively. $V_{ij}$ is the electrostatic Green's
function and $\delta n_i=n_i-n_i^0$ is the change in the
self-consistent charge $n_i$ from its value in zero source-drain
voltage $n_i^0$. This third term is the direct Coulomb
interaction created by the source-drain bias-induced charges. It
includes the Hartree term which is the solution of Poisson
equation and locates on the diagonal terms of Hamiltonian. For
calculating the electrostatic Green's function, we have used the
method explained in Ref.[\onlinecite{cheraghchi2}] and its
appendix.
\bfig
\includegraphics[width=10 cm]{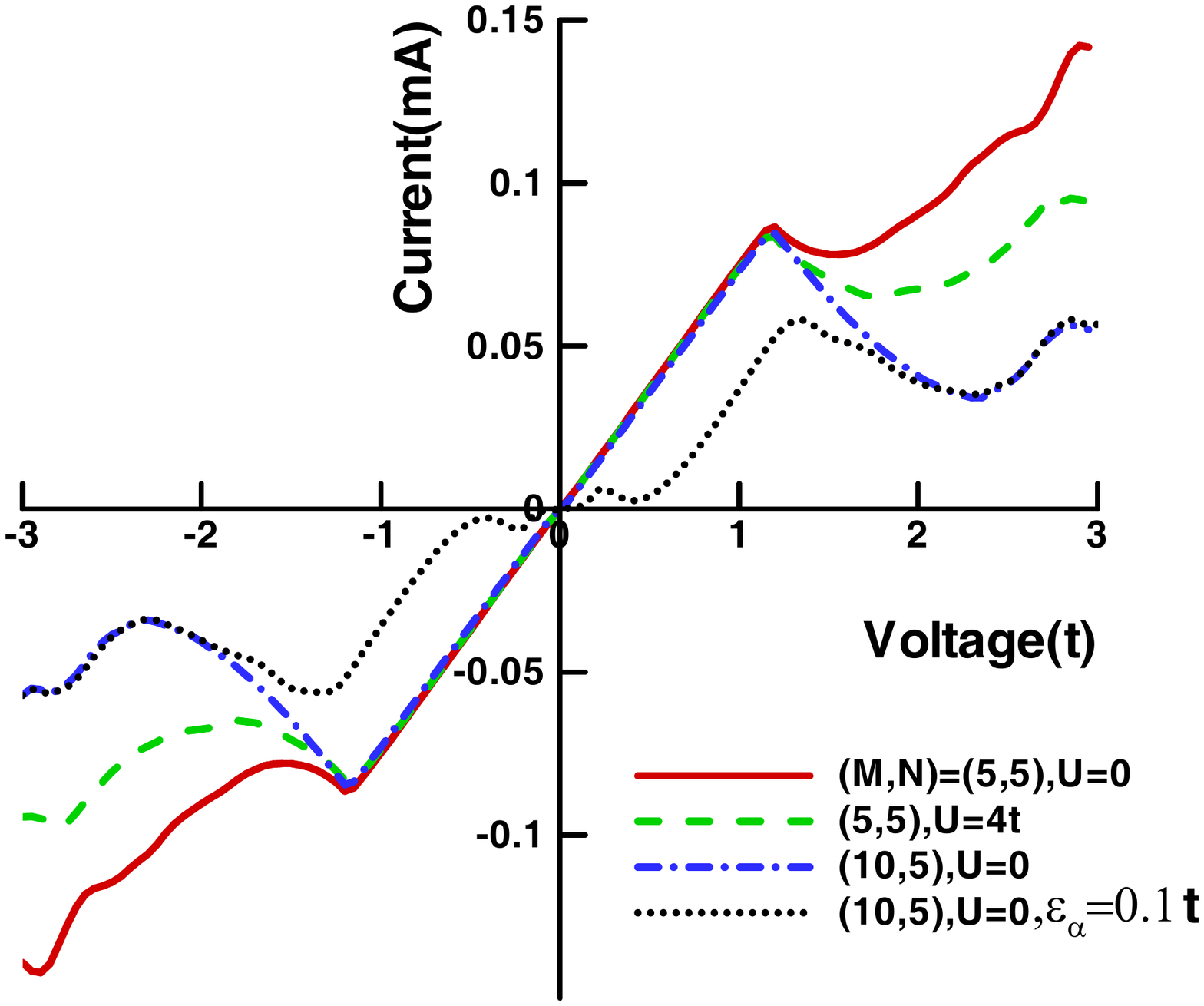}
\caption{Current-voltage characteristic curve of an odd zigzag
graphene nanoribbon with $N=5$ (zigzag chains). I-V curves are
compared for two ribbon lengths; $M=5$ and $10$. For the case of
$(M,N)=(5,5)$ and for comparison purpose, I-V curve is also
plotted in the presence of electron-electron interaction ($U$) and
asymmetric factor($\varepsilon_\alpha$).}\label{IV_oddZGNR} \efig
For self-containing, we present a very brief review of the NEGF
formalism. Charge density in non-equilibrium situation is
calculated by [$-iG^<$] as the occupation number in the presence
of the two electrodes with an applied source-drain
bias\cite{Taylor}.

\beq n_i=\frac{-1}{\pi}\int_{-\infty}^{E_F-\frac{V}{2}}
Im[G^{r}(E)]_{ii} dE+n^{non-eq}_i\label{charge} \eeq where
non-equilibrium part of charge can be calculated by the following
integral, \beq
n^{non-eq}_i=\frac{1}{2\pi}\int_{E_F-\frac{V}{2}}^{E_F+\frac{V}{2}}
[-iG^{<}(E)]_{ii} dE\label{noneq_charge}\eeq where within a
one-particle theory ,

 \beq -iG^<=G^r(\Gamma_L
f_L+\Gamma_R f_R)G^a \label{lesser_green}\eeq

Here $f_{L/R}$ is the Fermi-Dirac distribution function of
electrodes and $G^r/a$ is the retarded/advanced Green's function
defining as the following:
 \beq
G^{r/a}=[(E\pm\eta)I-H\{n\}-\Sigma_L^{r/a}-\Sigma_R^{ra}]^{-1}\label{Green}\eeq

and $\Gamma$ is the escaping rate of electrons to the electrodes
which is related to the self-energies as
$\Gamma_p=i[\Sigma_p^r-\Sigma_p^a]$ with
$p=L/R$\cite{Datta,Taylor,Munoz}. Here $\eta\rightarrow0^+$.
Solving equations \ref{charge} and \ref{Green} self-consistently
results in a self consistent charge and Green's functions.
Finally, the current passing through the molecule is calculated
by the Landauer formula for zero temperature \cite{Datta} which
is valid for coherent transport.

\beqnar I(V)=\frac{2e}{h}\int_{E_F-V/2}^{E_F+V/2}\,dE \,T(E,V)\
\label{current}\eeqnar

where $T(E,V)$ are the bias dependent transmission coefficient.

\beq T={\rm Tr}[G^r \Gamma_R G^a \Gamma_L]
\label{transmission}\eeq

Fig.(\ref{nanoribbon}) shows zigzag graphene nanoribbon with an
{\it odd} number of zigzag chains in width ($N$). Central
interacting region, left and right electrodes divide the ribbon
into three regions. Odd ZGNRs has a bisection plane which its
intersection with the ribbon has been shown as the x-axis in
Fig.(\ref{nanoribbon}). Although this plane bisects the ribbon, it
is not a reflection plane. If upper half of the ribbon is
displaced by $a/2$ in respect to the lower one, mirror symmetry
is achieved against x-axis. $a$ is the bond length of C-C. In
other words, part of the wave function which is a functional of
$y$ variable, is an eigenvector of the parity operator
\cite{Polini}.

\section{Results}

\bfig
\includegraphics[width=8 cm]{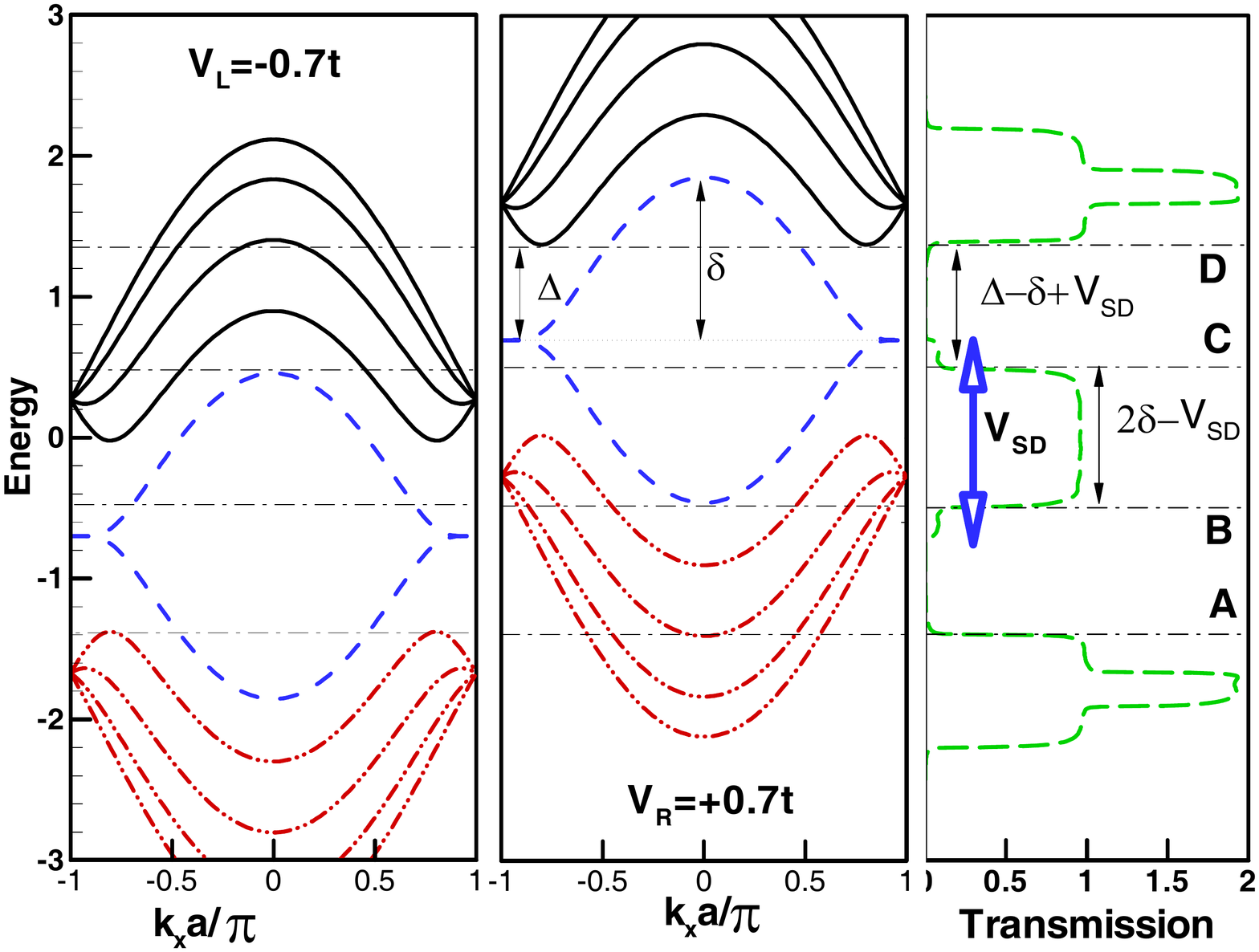}
\caption{Energy spectrum of the left and right electrodes and
transmission through zigzag graphene nanoribbon with
$(M,N)=(10,5)$ at voltage ($V_{SD}=1.4t>V_T$). The band structure
is divided into three groups which are called by upper , centeral
and lower band groups. These groups are classified based on the
bands which are connected in terms of $k_x$. The bold hollow
arrows show the current integration window which based on
Eq.~\ref{current}, is proportional to $V_{SD}$. The Fermi level is
set to be as $E_f=0$. The half-width of the central bands at
$k_x=0$ is called as $\delta$ which is equal to the threshold NDR
voltage. $\Delta$ is energy separation of the upper bands from
the central bands at the Dirac points. Transport gaps $AB$ and
$CD$ are equal to $\Delta-\delta+V_{SD}$. }\label{T_5layers} \efig
Current-voltage characteristic curve of an odd ZGNR with $5$
zigzag chains ($N=5$) is shown in Fig.(\ref{IV_oddZGNR}). Lower
than the external bias $1.2t$, current increases linearly with the
applied bias as an {\it Ohmic} device. After a threshold voltage
($1.2t$), NDR occurs at both positive and negative polarity.

The origin of NDR seen in odd ZGNRs is interpreted by analyzing
their energy spectrum accompanied to transmission curve.
Fig.(\ref{T_5layers}) shows energy spectrum $E(n,k_x)$ and
transmission through 5-ZGNR at $V=1.4 t$. $n$ counts bands from
the bottom ($n=1$) to the top of band structure ($n=2N$) and
$k_x$ is the longitudinal momentum.

In odd ZGNRs, parity has noncommutative relation with the
Hamiltonian. Therefore, parity has no conservation and
consequently transmission is not blocked by the parity selection
rule, while parity conservation in even zigzag nanoribbons opens
transmission gap around Fermi level\cite{cheraghchi1}. In the
range of $BC$ of Fig.(\ref{T_5layers}), there exists one
conducting channel which results in the unity transmission around
the Fermi level. In this range, electrons which occupy $-k$ states
of the lower band of the central band group (the dashed blue
bands) belonging to the right electrode are injected into the
unoccupied $-k$ states of the upper band of the central group
belonging to the left electrode. So at low biases, current
increases linearly with the bias. This single-channel transport
around the Fermi level remains unchanged even for high voltages.
However, for voltages greater than the NDR threshold voltage
($V>V_T$), blocked regions marked by the ranges of $AB$ and $CD$
comes into the current integration window. The current
integration window, based on Eq.~\ref{current} and $E_f=0$, is
proportional to $V_{SD}$ and is shown with the bold hollow arrows
in Fig.(\ref{T_5layers}). Therefore, when the source-drain applied
bias increases, current begins to decrease.

Blocked regions ($AB$ and $CD$) arise from a selection rule which
increases back scattering in the lengthy ribbons. According to
this rule, electron transition between those bands which are
disconnected from the view point of longitudinal momentum,
decreases exponentially with the length. Topology of zigzag
graphene ribbons divides the band structure into the three
different groups\cite{grosso}. Three groups which are called by
upper (n=1,2,3,4), centeral (n=5,6) and lower (n=7,8,9,10) groups,
are classified based on the bands which are connected in terms of
$k_x$. Variation of the momentum of electrons passing through the
system strongly depends on smoothness or sharpness of the
potential. The transition probability of electrons with the $k$
state injected by the right electrode into the $q$ state as an
empty state in the left electrode is proportional to the Fourier
transform of the longitudinal voltage \cite{grosso} as
$<\psi(k)\mid V \mid \psi(q)>\propto \tilde{V}(k-q)$. So smooth
variation of the potential in longer ribbons results in a small
momentum variation of electron. Consequently, a smooth potential
in the longitudinal direction can just scatter electron into
those states belonging to the energy bands which are connected
from the point of momentum view\cite{grosso,cheraghchi1}.
\bfig
\includegraphics[width=8 cm]{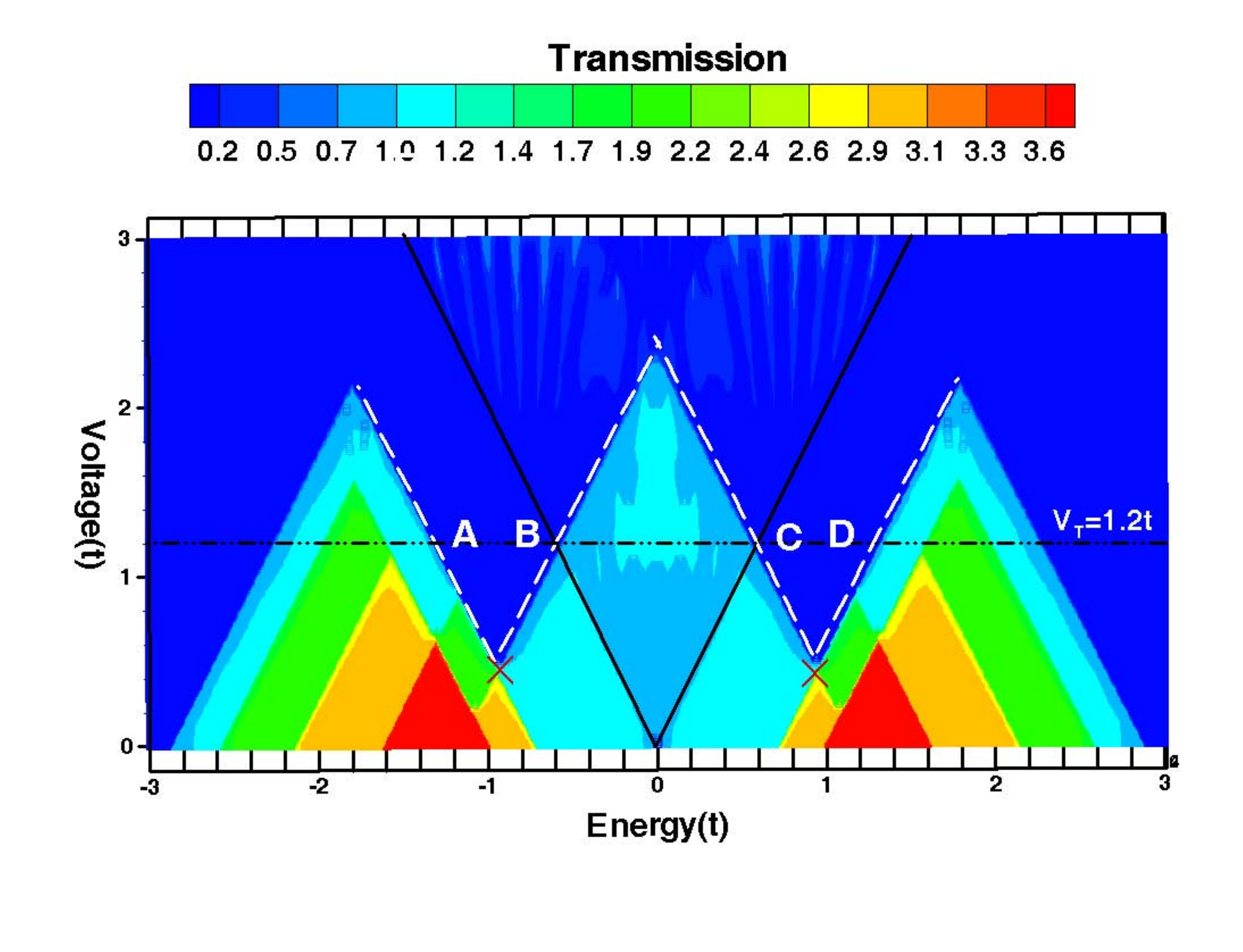}
\caption{Contour plot of transmission versus energy and applied
bias for graphene nanoribbon with 5 zigzag chains in width and 10
unit cells in length. Dark solid lines show the current
integration window and white dashed lines show the forbidden
region for electronic transition from the band groups of upper or
lower bands to the bands of central group.}\label{3Dtransmit}
\efig

Bias dependent of transmission leads us to plot a contour of
transmission in plane of energy and applied bias as shown in
Fig.(\ref{3Dtransmit}). In this figure, the solid lines show the
current integration window. Moreover, the dashed lines clarify
the regions $AB$ and $CD$ corresponding to transport gaps shown in
Fig.(\ref{T_5layers}). The intersection of these blocked regions
and also the current integration window would be around $1.2t$
which is the threshold voltage $V_T$ for the current reduction.

Band structure analyzing demonstrates that the threshold voltage
is equal to the half-width of the central bands at $k_x=0$ as
$V_T=\delta=[E(N+1,k_x=0)-E(N,k_x=0)]/2$. As shown in
Fig.(\ref{T_5layers}), $\Delta$ is energy separation of the upper
bands from the central bands at the Dirac points. There is a
Log-Normal behavior of $\delta$ versus number of zigzag chains
($N$) such that as $N\rightarrow\infty$, the NDR threshold
voltage asymptotically approaches to the value of
$0.9738\pm0.0002 t$. So the NDR threshold voltage slightly
decreases with the ribbon width.

Analyzing transport gaps appeared in the band structure shows that
they are equal to $\Delta-\delta+V_{SD}$ where $\Delta \propto
N^{-1}$. Since $\delta$ approaches to a constant values when
$N\rightarrow\infty$, in a given voltage, transport gap is
disappeared for $N>30$ which is nearly equivalent to $10 nm$.

The other effect which enhances performance of this electronic
switch, is the ribbon length. Fig.(\ref{IV_oddZGNR}) shows an
increase of on/off ratio with the ribbon length. Moreover, NDR
region ($V_{off}-V_{on}$) occurs in a more extended range of the
I-V curve. Exponential decay of transmission with the length in
the gap regions develops quality of switching. It is shown in
Fig.(\ref{onoff}) that $I_{on}/I_{off}$ increases exponentially
with the ribbon length.

\bfig
\includegraphics [width=8 cm]{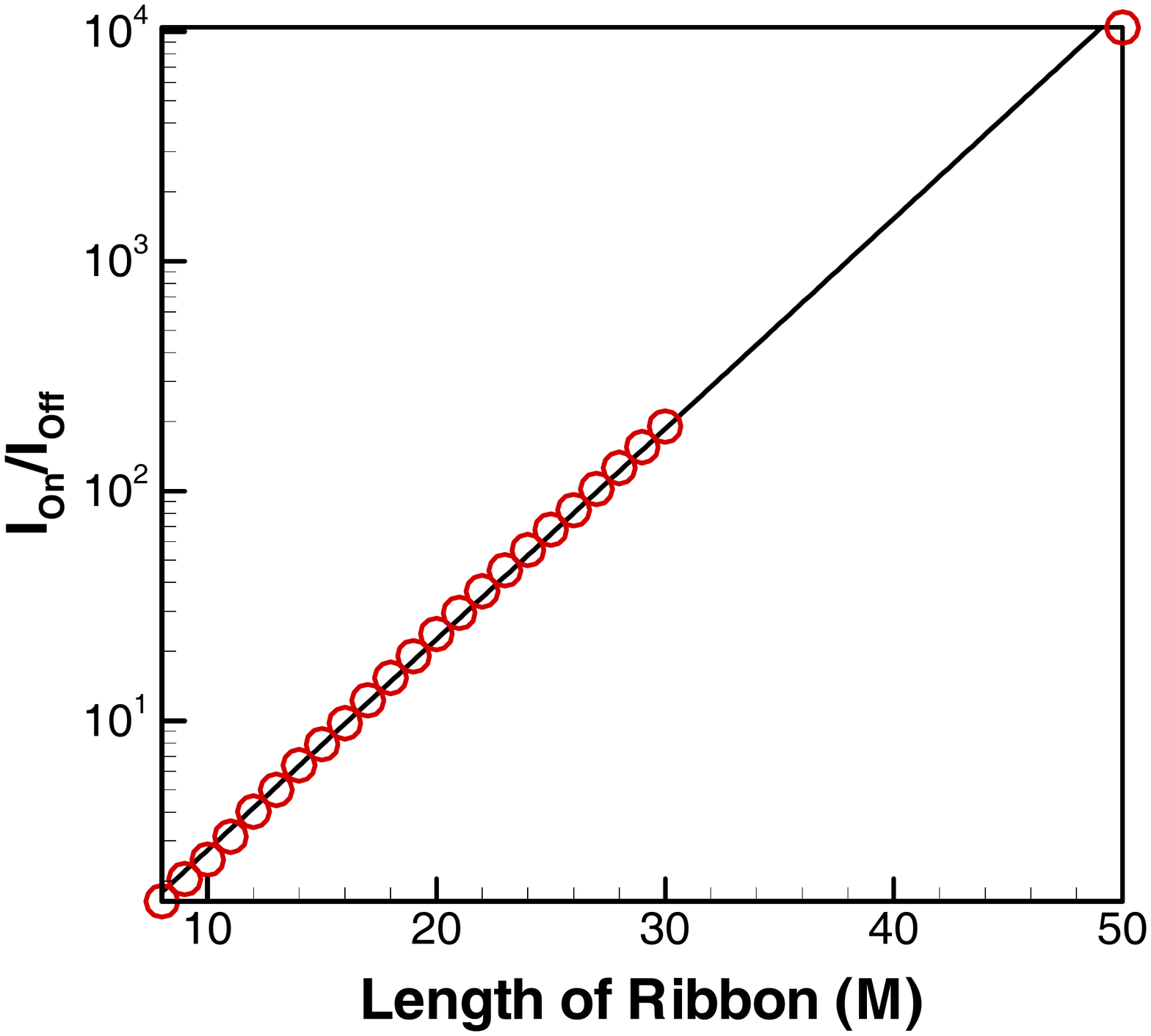}
\caption{Exponential increase of on/off current ratio as a
function of ribbon length ($M$).\label{onoff} }\efig
Another parameter which affects I-V curve, is the e-e interaction.
In this case, interaction intensifies NDR effects such that off
current reduces in compared with non-interacting system and also
the region of NDR becomes more extended.

If one of the ribbon edges is doped by small impurity such as
$\varepsilon_\alpha=0.1t$, because of a band gap which is induced
by edge impurity at the band center, current at low biases
decreases. Fig.(\ref{IV_oddZGNR}) represents that even with the
presence of edge impurity, still the region containing NDR exists.
However, asymmetry decreases on/off ratio of the current.
Furthermore, asymmetric ZGNR behaves as a semiconductor while
symmetric ZGNRs behave as an Ohmic devices\cite{Ren}. The effect
of asymmetry on NDR competes with the ribbon length. Since
asymmetry can not be ignored in experiment, longer ribbons are in
favor of keeping NDR in the I-V curve.

On the other hand, additional to edge impurity, this asymmetry can
be assigned to a sublattice symmetry breaking induced by
spontaneous ferromagnetic spin ordering of the electrons
localized at the zigzag edges\cite{Louie}. In fact, the border
atoms at the two opposite zigzag edges belong to different
sublattices. So spin orientation along the edges induces
different magnetic potentials at the edges. As a result, a small
band gap is opened around Fermi level which depending on the
ribbon width, is about 0.15 eV. The asymmetry which we have
considered is about 0.3 eV which is stronger than the gap opened
by spin-polarization of the edges. We can conclude that
spin-polarization along the zigzag edges can not affect emerging
of this NDR phenomenon\cite{Ren}.

{\it Charging effects}: To understand why details of electrostatic
potential do not affect the emergence of NDR, we compare
transmission in the presence and absence of e-e interaction.
Figure (\ref{int-nonint-T}) compares transmission curves for
voltages $1.2t$ at the NDR threshold $V_{T}$ and a larger value
such as $1.5t$. It is clear that for voltages lower than the NDR
threshold, transmission in the conducting region $BC$ is robust
against e-e interaction. However, for voltages $V>V_{T}$,
transmission of conducting region $BC$ reduces with e-e
interaction. This is the reason for reducing off-current in the
presence of e-e interaction. Furthermore, transmission curve in
the blocked regions $CD$ and $AB$ exhibits a little enhancement
with the e-e interaction for the whole range of voltages.

\bfig
\includegraphics[width=8 cm]{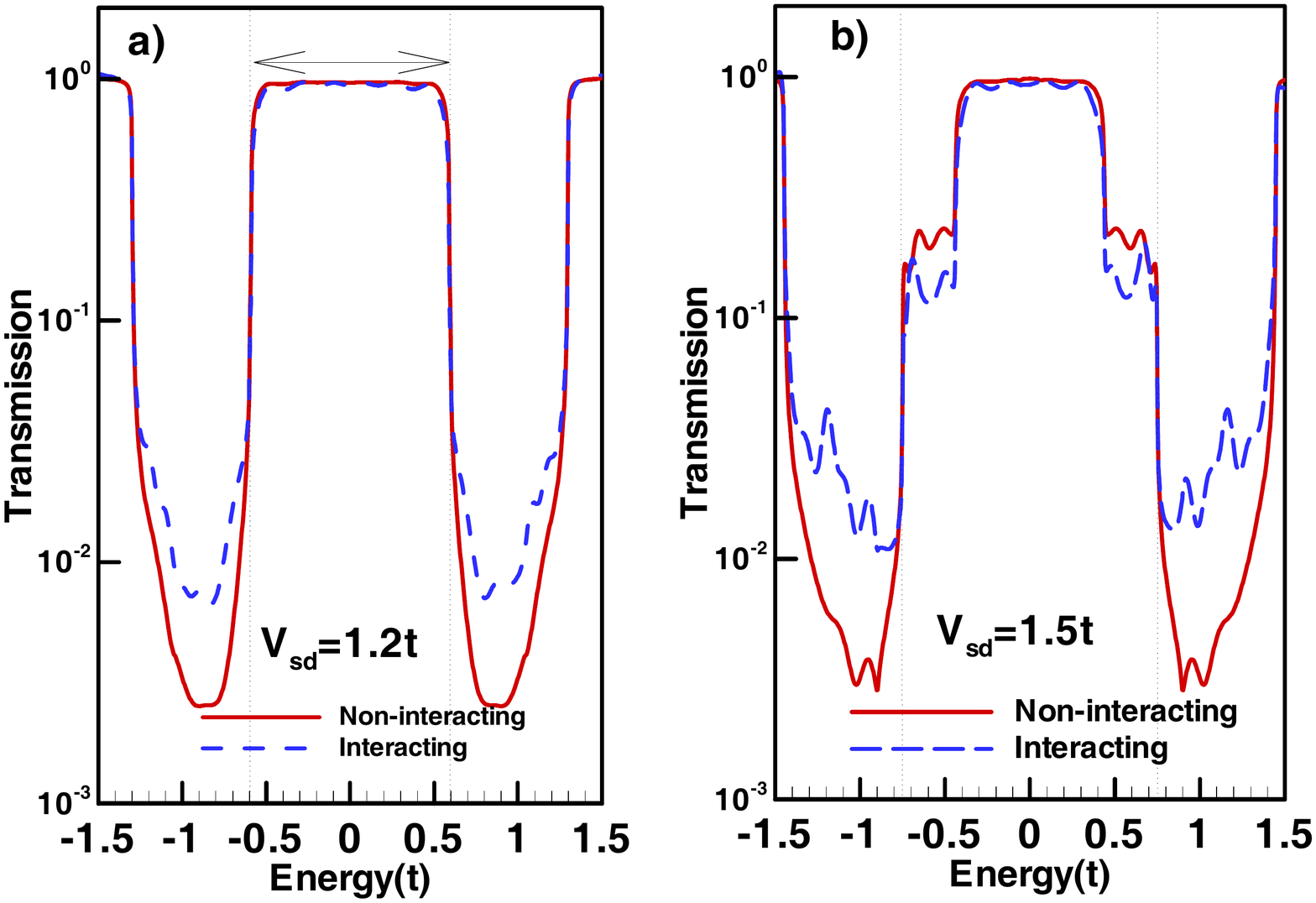}
\caption{The effect of electron-electron interaction in the
mean-field level on transmission curves for two voltages a) NDR
threshold voltage which is equal to 1.2t, b) a voltage above the
NDR threshold such as 1.5t.}\label{int-nonint-T} \efig
To elucidate physics behind to the robustness of NDR phenomenon
against e-e interaction, we investigate self-consistent potential
profiles which are shown in Fig.(\ref{SCF-charge-potential}.a,b).
In fact, external potential is well screened by charge
redistribution so that potential mostly drops at the contact
junctions. This fact is represented in
Fig.(\ref{SCF-charge-potential}). So, electrostatic potential of
atoms located far from contact junctions (deep inside the ribbon)
remains nearly flat. In more details, depletion of charge close
to the source electrode reduces the source potential as $U \delta
n$. Therefore, source potential can not penetrate inside the
central portion of the system. On the opposite side of system,
charge is accumulated close to the drain electrode such that
dropping of the potential is weakened.
\bfig
\includegraphics[width=8 cm]{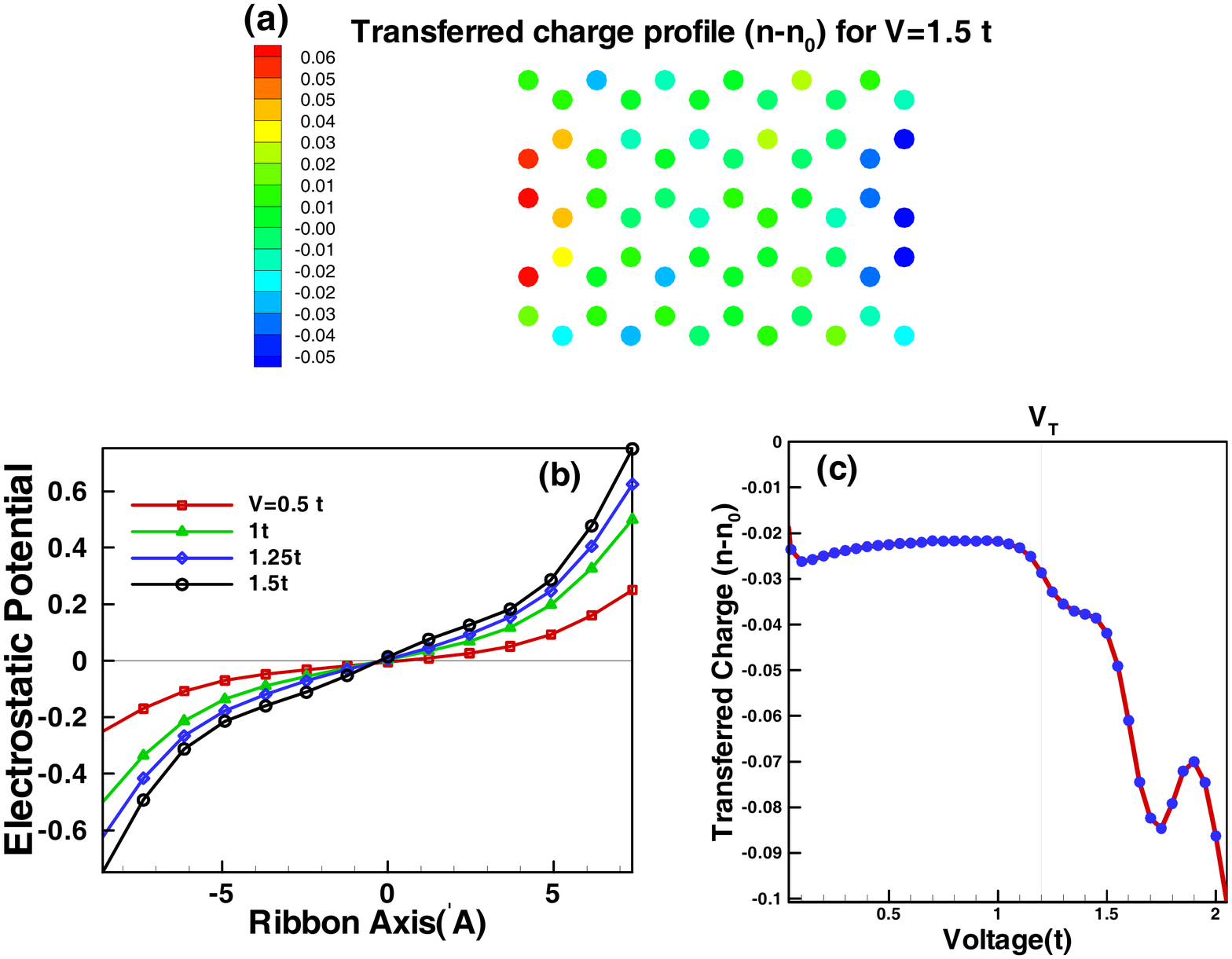}
\caption{a) Self-consistent charge profile at 1.5t. Charge
compares with its neutrality point ($n_0$) when there is no
applied bias. Transferred charge is equal to $(n-n_0)$.
Electrostatic potential profile is calculated according this
transferred charge. b) averaged electrostatic potential variation
on unit cell as a function of the ribbon axis.  c) Total
transferred charge from/into the ribbon as a function of
source-drain voltage. }\label{SCF-charge-potential} \efig
It is useful to investigate bias dependence of total transferred
charge from/into the system. Fig.(\ref{SCF-charge-potential}.c)
shows that transferred charge in compared with its neutrality
point value ($n_0$ with zero source-drain applied bias) decreases
for voltages higher than the NDR threshold, while it remains
unchanged for voltages $V<V_{T}$. In other words, for $V<V_{T}$,
in-flowing and out-flowing charges are balanced with each other,
while for $V>V_{T}$, system becomes empty of charge. It can be
seen that the threshold voltage for depletion of charge is
correlated with the NDR threshold voltage. This behavior is
closely similar to what is seen in graphene
nano-junctions\cite{cheraghchi2}. As a consequence, at high
voltages, due to accumulation of charge close to the drain
electrode, external potential is always screened in the middle of
the graphene strip. Screening effect induces sharp variation of
electrostatic potential at the contacts which enhances transition
probability between disconnected band groups. Therefore,
transmission in the blocked regions ($CD$ and $AB$) increases in
compared to non-interacting system. However, blocked regions have
not any contribution in the current calculation.

\bfig
\includegraphics[width=8 cm]{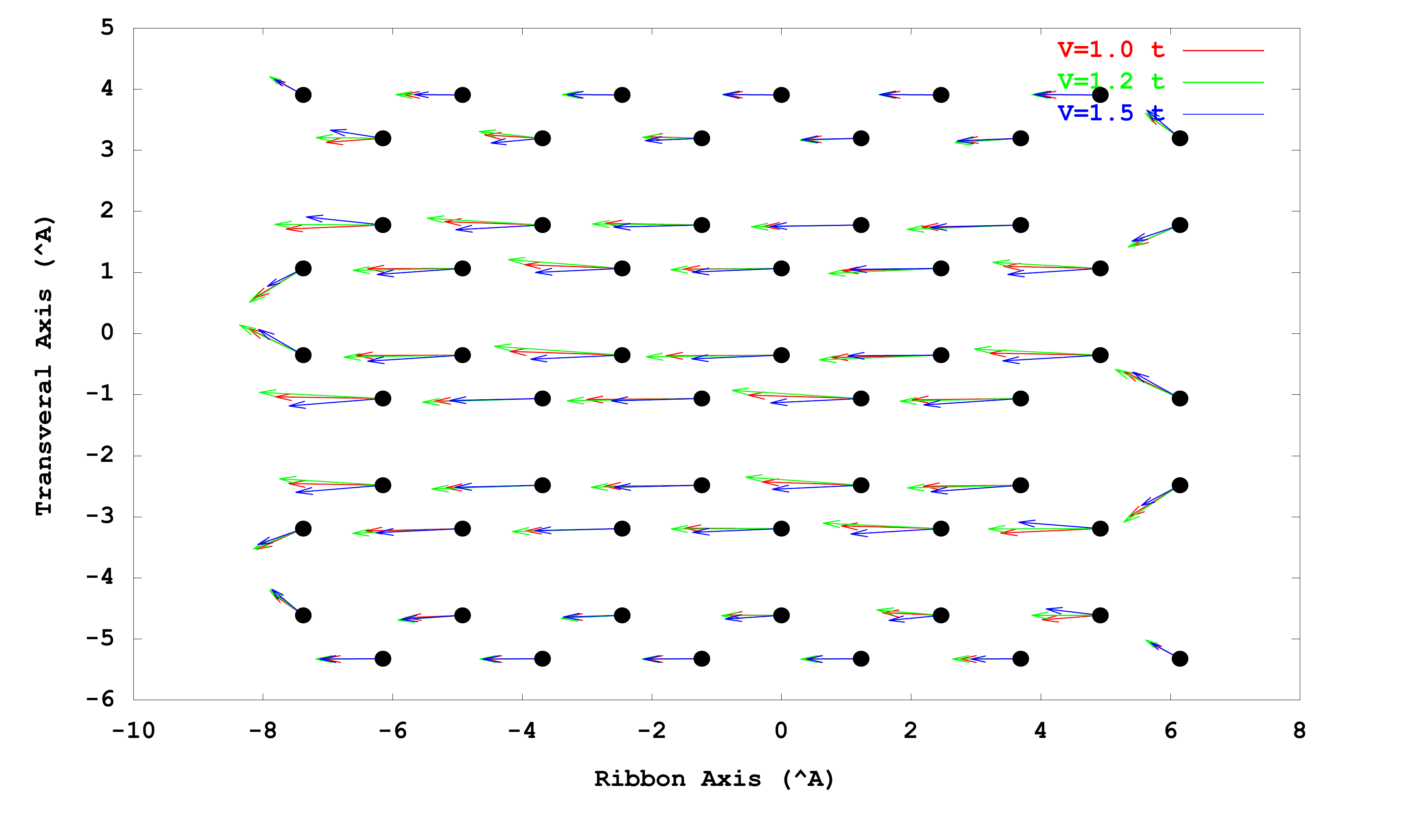}
\includegraphics[width=8 cm]{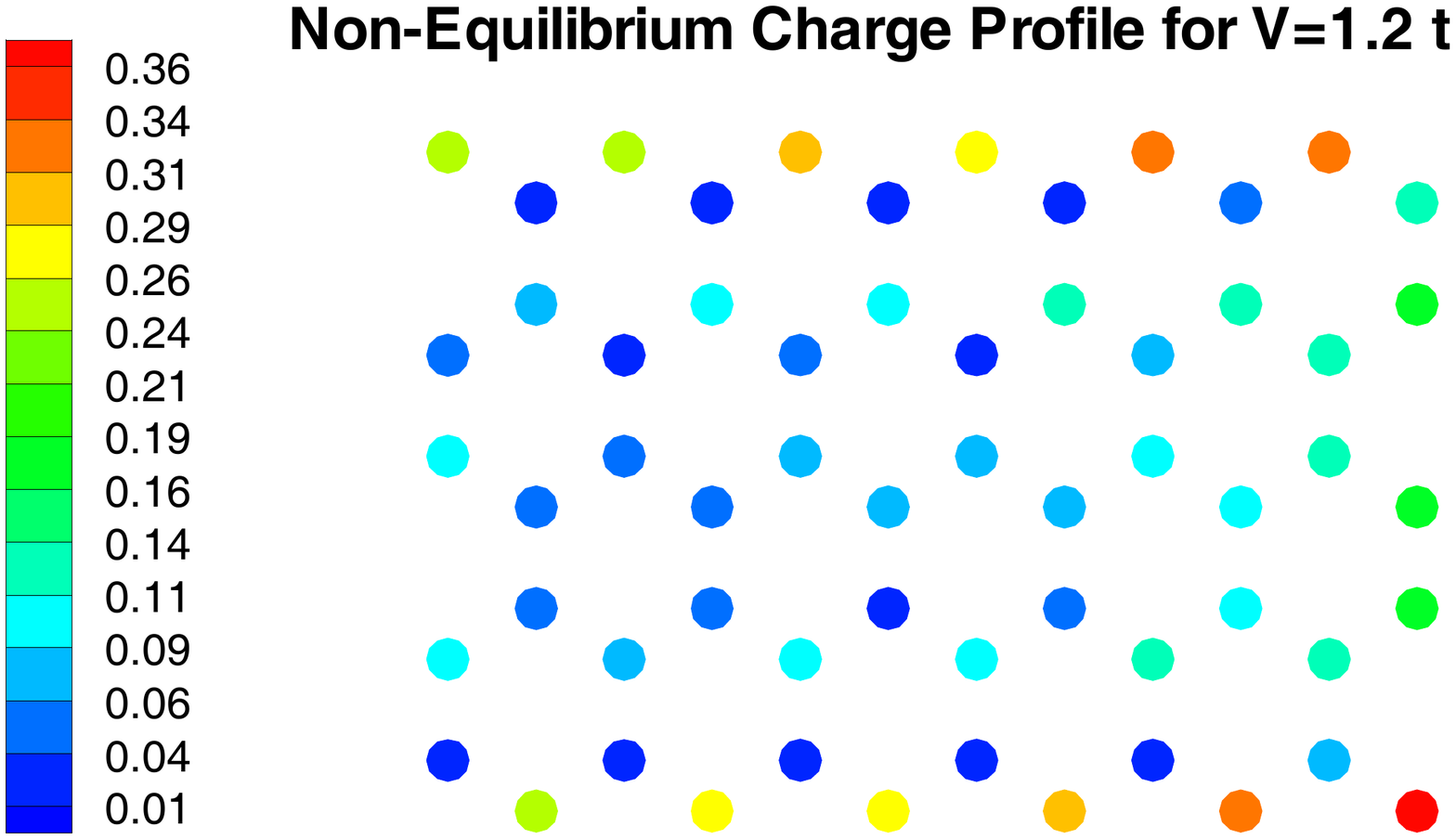}
\caption{Top: Spatial distribution of bonds current in three
voltages: red arrows) 1.0 t$<V_T$, black arrows) 1.2t $=V_T$ and
blue arrows) 1.5 t$>V_T$. The length of arrows is proportional to
the local current. The arrows located on the edges are shorter
than the arrows located in the bulk of nanoribbon. By increasing
the source-drain voltage, for voltages lower than $V_T$, current
increases and consequently arrows are longer, while higher
than $V_T$, current decreases and so arrows are shorter. \\
Bottom: Non-equilibrium charge profile at the threshold NDR
voltage. Non-equilibrium charge reaches in its maximum value at
the ribbon's edges. Moreover, in agreement with the continuity
equation, atomic sites with larger local current corresponds to
smaller non-equilibrium charge.}\label{local_current} \efig

{\it Spatial distribution of bond currents}: Charge conservation
based on continuity equation yields non-equilibrium bond charge
current\cite{nikolic} for a non-interacting tight-binding
Hamiltonian.

\beq J_{ij}=\frac{2et}{h}\int_{E_F-\frac{V}{2}}^{E_F+\frac{V}{2}}
[G_{ij}^{<}(E)-G_{ji}^{<}(E)] dE\label{bond_current}\eeq where
sites $i$ and $j$ are nearest-neighbor atoms which their hopping
integral is nonzero. Now, charge continuity equation is derived
by using Heisenberg equation.

\beq e\frac{d}{dt}n^{non-eq}_i+\sum_j[J_{ij}-J_{ji}]=0
\label{continuity_equation}\eeq where $j$ are the
nearest-neighbor atomic site around $i^{th}$ atomic site. It can
be simply proved that the integrands in Equations
\ref{noneq_charge} and \ref{bond_current} are real. In fact,
transpose of matrix $-iG^<$ in Eq.(\ref{lesser_green}) is equal
to its conjugate. So, diagonal terms of ($-iG^<_{ii}$) and also
the terms of ($G^<_{ij}-G^<_{ji}$) are real.

Bond charge current formulated in Eq.\ref{bond_current} has been
derived for non-interacting tight-binding Hamiltonian. However,
by using Heisenberg equation, it can be simply proved that this
formula is still applicable for the Hamiltonian described in
Eq.\ref{hamiltonian} in which Hartree interaction appears on
diagonal elements of the Hamiltonian. On the other hand, charging
effects originating from electrostatic interaction (Hartree term)
adjust local currents by means of lesser Green's function.

Fig.(\ref{local_current}.top) shows spatial profile of local
current at each site. In the center of the ribbon, magnitude of
local current densities are larger than its value in the zigzag
edges of the ribbon. This feature is valid for both ranges of low
and high voltages not more than 1.5 t. The reason can be followed
by single-channel transport through ZGNR even for high
voltages\cite{nikolic}. One channel transport is an indirect
result of transition rule in which transition between
disconnected bands is forbidden. For voltages higher than 1.5 t,
alignment of local currents along the ribbon axis is gradually
disturbed.
\bfig
\includegraphics[width=8 cm]{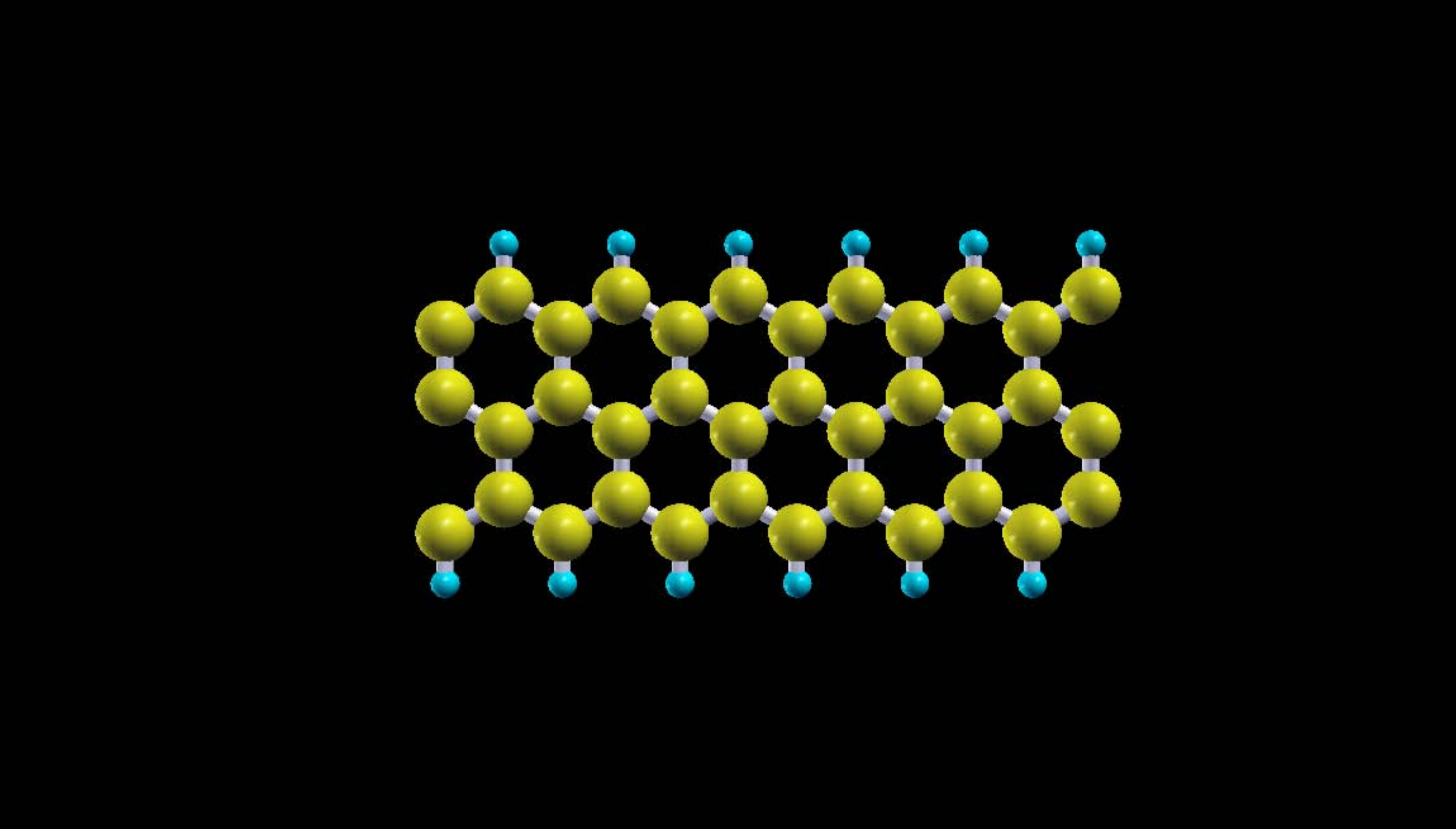}
\includegraphics[width=8 cm]{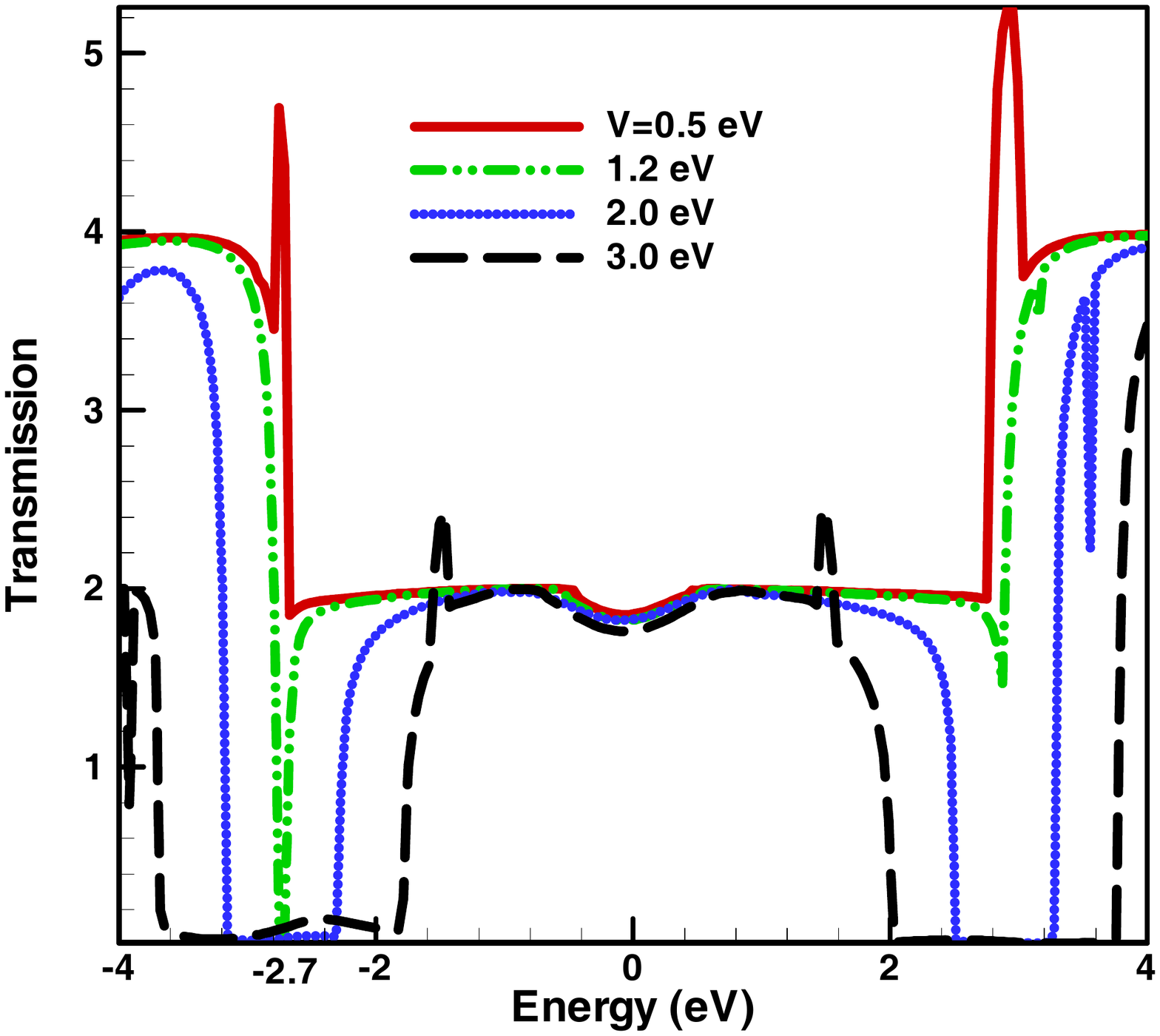}

\caption{Top: Zigzag graphene nanoribbon terminated by Hydrogen
atoms. Bottom: First principle calculation of transmission through
the above system as a function of energy for different values of
external biases. }\label{terminated_GNR} \efig

In contrast with the local current distribution, as shown in
Fig.(\ref{local_current}.bottom), non-equilibrium charge reaches
in its maximum value at the ribbon's edges. This result is
compatible with the continuity equation
(Eq.\ref{continuity_equation}) so that atomic sites with larger
local current corresponds to smaller non-equilibrium charge. On
the other hand, non-equilibrium charge increases for atomic sites
with higher onsite energy due to applied bias (close to the source
electrode), while its value tends to zero for sites close to the
drain electrode.

{\it Comparison of odd and even ZGNRs}: There are some
interesting differences between results arising from {\it odd}
ZGNRs with those results belonging to {\it even}
ZGNRs\cite{cheraghchi1} which is kind of odd-even effect. Here,
NDR appears in voltages upper than $1t$ while in even ZGNR, NDR
occurs for voltages lower than $1t$. On/off ratio of the current
in gated even ZGNRs increases as a power law with the function of
the ribbon length while here, on/off ratio increases
exponentially. Screening of the external bias by electrons of
system in even ZGNRs is so stronger than screening effects in odd
ZGNRs. As a consequence, the effect of electrostatic interaction
on increase of the on/off ratio in even ZGNRs is much effective
than in odd ZGNRs. Furthermore, transferred charge from/into the
central portion of graphene nanoribbon depends on odd or even
zigzag chains in width.

\section{Hydrogen Terminated Nanoribbon}
To show how passivation of zigzag edges of graphene nanoribbon
affects transport properties, we compare the results of the
presented model with transport properties calculated by using
TranSIESTA code\cite{siesta}. This code is based on density
functional theory (DFT) approach. We have used the following
options to calculate SIESTA code: the generalized gradient
approximation (GGA) with the Perdew-Burke-Ernzerhof
exchange-correlation functional (PBE)\cite{PBE}, double-$\zeta$
plus polarization orbital (DZP) bases for all atoms, and
Troullier-Martins norm conserving pseudopotentials to represent
the cores,  200 Ry real space mesh cutoff for charge density and
a supercell within 20 $\AA$ of vacuum between the periodic
graphene nanoribbons.

Fig.(\ref{terminated_GNR}, Top) shows a schematic view of odd
zigzag graphene nanoribbon which is saturated by hydrogen atoms.
Transmission around the Fermi energy is represented in Fig.
(\ref{terminated_GNR}, bottom). The energy scale is shifted so
that the Fermi energy of the system, if there is no bias voltage,
is zero ($E_F=0$). When there is a finite bias, the Fermi energy
of the left electrode is placed at $V/2$ and the right electrode
at $-V/2$. The occupation number on atoms in the center part of
the system is determined by $[-iG^<]_{ii}$.

As it is clear of Fig.(\ref{terminated_GNR}, bottom), there exists
one transmitting channel for each spin around Fermi energy level.
This result is in complete agreement with the band structure
analysis shown in section.III. There exists only one transmitting
channel in the region $BC$ presented in figures \ref{T_5layers}
and \ref{3Dtransmit}. The transport gap in Fig.
(\ref{terminated_GNR}, bottom) begins to open at voltage around
$1.2 eV \simeq 0.5t$ and for energy about $2.7 eV \simeq t$. These
points correspond to the points marked by cross sign in
Fig.\ref{3Dtransmit} in which the transport gap begins to open as
the applied bias increases. As a result, we demonstrate that the
transport gap which is responsible for emerging the NDR phenomena
could also be opened in graphene nanoribbons with passivated
zigzag edges.

\section{Conclusion}
As a conclusion, based on a model calculation of non-equilibrium
Green's function formalism, we found that there exists a region
of negative differential resistance in I-V curve of ultra narrow
(lower than $10nm$) zigzag graphene nanoribbons with odd number
of zigzag chains in width. This NDR is induced by a transport gap
which originates from electronic transition between disconnecting
bands of energy from the view point of longitudinal momentum.
On/off ratio of the current exponentially increases up to $10^5$
as a function the ribbon length which proposes possibility of
manipulation of odd ZGNRs as high quality switch in
nanoelectronic based on graphene nanoribbons. In addition, e-e
interaction enhances on-off ratio of the current which originates
from a flat electrostatic potential deep inside the ribbon due to
screening of the external bias by electrons close to the
junctions. By using continuity equation, spatial profile of local
currents is calculated in the presence of Hartree electron
interactions. In both high and low biases, local current reaches
to its maximum values in the center of the ribbon while in
contrast with the local current profile, non-equilibrium charge
has its maximum values at the edges of the ribbon. Furthermore,
this NDR is not much sensitive to the edge asymmetry. So emerging
of this NDR is robust against spin orientation along the edges.

By using {\it ab initio} density functional theory, we also show
that the transport gap which is responsible for emerging NDR,
exits in passivated graphene nanoribbons by Hydrogen atoms.

\end{document}